# AI Surge Accelerates the Transition to Liquid Cooling: Rethinking Energy and Water Use in Data Center Cooling

Yingbo Zhang[1,2], Eric Masanet[2,3], Shengwei Wang[1,4*]

[1] Department of Building Environment and Energy Engineering, The Hong Kong Polytechnic University, Hong Kong

[2] Bren School of Environmental Science and Management, University of California, Santa Barbara, CA, USA

[3] Department of Mechanical Engineering, University of California, Santa Barbara, CA, USA

[4] Research Institute for Smart Energy, The Hong Kong Polytechnic University, Hong Kong

*Correspondence: beswwang@polyu.edu.hk

## Abstract

The power density of AI workloads is rapidly driving shifts to liquid cooling. This study rethinks and assesses the global impact on energy and water demand of this cooling transformation. Results suggest that liquid cooling can reduce cooling energy use by 25-56% relative to air cooling, offering substantial energy savings for GW-scale AI data centers. Best-practice PUE values could be as low as 1.07-1.08 for today's systems and 1.03-1.04 for emerging designs and WUE values from 0.38 to 2.63 L $kWh^{-1}$, depending on temperature conditions. We identify global free-cooling rack inlet temperature thresholds of 33 °C for evaporative systems and 49 °C for dry systems, demonstrating the practicality of 100% free cooling in many locations. The model finds that minimum PUEs do not necessarily coincide with minimum total energy use, because warmer coolant raises server power. Results quantify the large sustainability potential of liquid cooling and provide actionable targets for data center design and policies.

## INTRODUCTION

The world is currently experiencing an unprecedented technological revolution, commonly referred to as the AI boom [1], that is reshaping industries and infrastructure on a global scale. The release of ChatGPT in late 2022 marked a major milestone in artificial intelligence (AI), triggering rapid development of increasingly capable large language models. Serving as the backbone of high-performance computing, data centers have played a pivotal role in driving these advances and are expanding rapidly worldwide [1]. According to a recent IEA report, data centers have been a major driver of global electricity demand growth in recent years [2]. China and the United States are the most significant regions for data center electricity consumption growth, accounting for nearly 80% of the estimated global growth to 2030 [1]. The dramatic increase in the energy demands of AI data centers worldwide leads to high-density heat generation, making cooling technologies one of the most pressing challenges of AI data centers.

Cooling technologies in data centers have evolved considerably over the past few decades [3]. Early internet-era facilities relied on adaptations of conventional air conditioning systems, leading to the development of computer room air conditioners tailored for server environments [4]. With the expansion of internet infrastructure and the rise of social media and smartphones, cooling strategies began focusing on airflow management to improve efficiency, especially with the widespread adoption of aisle containment systems [5]. In recent years, advances in AI-assisted temperature control have further refined airflow management [6]. Additionally, efforts to raise ambient temperature thresholds within server rooms have enabled the wider use of free cooling, a technique that leverages cool external air to reduce mechanical cooling loads. However, despite these improvements, air cooling of racks is struggling to cope with the ever-growing heat flux from high-performance computing chips. Rack density of next-generation AI data centers is estimated to reach up to 130 kW/rack [7], representing a nearly 19-fold increase over 2021 levels [8]. In the longer term, projections of AI rack power densities of 500-600 kW/rack are increasingly common [9,10].

Ever-growing heat dissipation densities of new generation chips, such as GPUs, are now pushing liquid cooling toward mainstream deployment, especially in high-end AI training clusters [11]. Liquid cooling offers superior thermal performance through much higher heat capacities compared to air by bringing the coolant closer to the heat sources. Among existing approaches, cold-plate cooling has proven its reliability and engineering maturity through extensive deployment. At the research frontier, a recent study demonstrates that a microfluidic cooling strategy can dissipate heat fluxes up to 3,000 W/cm$^2$ at a pumping power of only 0.9

W/cm$^2$ using single-phase water as the coolant [12]. In parallel, a 3D-printed hierarchical micro/nano-structured surface has achieved a record-high spray cooling performance, reaching a heat transfer coefficient of 443.7 kW/(m$^2$·K) over a 1 cm$^2$ heating area [13]. Beyond laboratory-scale demonstrations, state-of-the-art industrial technologies have also adopted microchannel cooling in their new generation products. Notably, NVIDIA's Vera Rubin chip employs a microchannel-based cooling architecture that enables effective thermal management using warm water at rack inlet temperatures up to 45 °C [14]. Commercial cooling solutions like Frore Systems' LiquidJet Nexus claim to support next-generation AI accelerators (e.g., NVIDIA Blackwell) while reportedly maintaining operation with 53°C inlet coolant temperatures [15].

Another critical challenge associated with data center cooling is onsite water consumption. While the offsite water consumption associated with electricity generation, often referred to as "indirect" water, is also a critical challenge [16,17], it is beyond the scope of the current paper. Onsite water sourcing has become a pressing issue that many organizations operating data centers must confront directly [18]. To achieve higher cooling efficiency, many data centers have historically adopted evaporative cooling, such as cooling towers, [3], where water is evaporated into the atmosphere to dissipate heat, or adiabatic systems, where water evaporation is used to cool incoming air streams. These technologies improve cooling system energy efficiency, but they can lead to significant water loss and growing environmental concerns. While the onsite water use of data centers may seem small at global levels or compared to water-intensive sectors like agriculture, it can be significant at local scales in water-stressed regions [19]. Notably, nearly 75% of the global population now lives in water-insecure or critically water-insecure nations [20]. Data centers can consume large volumes of water directly for cooling, in some cases, the majority of which is sourced from potable water [21]. A recent U.S. report indicated that the rapid expansion of data centers might deplete the finite water supplies available to residents of Phoenix and Tucson, as Arizona's water resources are drying up [22]. Google's data centers in The Dalles consumed 355 million gallons of water in 2021, representing 29% of the city's total water consumption, which has raised concerns about local water stress [23]. Microsoft's global onsite water use soared by 34% while it was developing its initial AI tools, and a data center cluster in Iowa used 6% of the district's water supply in one month during the training of OpenAI's GPT-4 [24].

The ongoing evolution of data center cooling, driven by the AI revolution, is poised to reshape the energy and water demands of the global digital infrastructure. This study provides a novel global evaluation of the energy and onsite water impacts, as well as the associated trade-offs, of the AI-driven transformation of data center cooling systems. Using typical meteorological year data across 57 representative global cities covering 19 climate zones [25] (see Table S1) and alignment with major global data center hubs, we simulate the operation of state-of-the-art liquid cooling systems and compare them with conventional air-cooled configurations. Our results show that liquid cooling could reduce cooling energy consumption by 25–56% relative to air cooling under typical rack inlet temperature conditions. Under best-practice conditions, total Power Usage Effectiveness (PUE) values can achieve as low as 1.07-1.08 for today's liquid cooling systems and 1.03-1.04 for emerging high-performance liquid cooling designs if coupled with 99% efficient UPS and 99.5% power distribution systems [16]. We also identify globally applicable rack inlet water temperature thresholds that enable nearly 100% free cooling, where either cooling towers or dry coolers handle the cooling load without the need for mechanical chillers, across all climate zones. Moreover, we found that the temperatures corresponding to the minimum total system energy consumption and the minimum PUE do not necessarily coincide, and that the temperatures minimizing total system energy consumption are climate-dependent. By accounting for the impact of elevated inlet water temperatures on GPU server power consumption, we further identified the specific temperature that minimizes total system energy consumption in each climate zone. Additionally, the study discusses the energy performance and onsite water-use trade-offs between evaporative-cooled and dry-cooled liquid cooling systems (see schematics in Figure S1). We further evaluate the robustness of these findings across climates and operating conditions. These findings can be used to inform design and operational decisions for future AI data centers and to support long-term planning and policy analysis by quantifying climate-dependent energy and water trade-offs of advanced liquid cooling technologies.

## RESULTS

### Energy Performance and Free Cooling Potential of Liquid Cooling Systems

The energy performance of liquid cooling systems is strongly influenced by the rack inlet water temperature and local climatic conditions. These factors determine the availability of free cooling hours, which is defined as the periods when outdoor ambient air can be used directly for cooling without mechanical refrigeration (i.e., chiller operation). The ideal scenario is to achieve 100% free cooling throughout the year. According to ASHRAE standards [26] and manufacturer specifications [27], the typical inlet water temperature for server liquid cooling systems ranges from 20°C to 45°C. However, engineering applications have demonstrated the feasibility of operating at temperatures as high as 50°C [28,29]. Therefore, a comparative investigation is conducted to evaluate the energy performance of commonly used cold-plate liquid cooling systems across an extended temperature range (20°C to 50°C), considering both commonly used dry-cooled and evaporative-cooled systems for building-level heat rejection (see Figure S1). Using weather data [30] from 57 representative cities across 19 global climate zones [25], the free cooling potential of both systems is evaluated for inlet water temperatures between 20 °C and 50 °C.

Figure 1 illustrates the free cooling potential and the Cooling Power Usage Effectiveness (CPUE), a partial PUE metric [31] used to isolate the cooling system's contribution to the overall PUE, for both dry-cooled and evaporative-cooled systems across various climate zones and rack inlet water temperatures. Figures 1a and 1b present heatmaps of the annual free cooling ratio, which is defined as the number of hours per year during which free cooling is available divided by 8760 hours (i.e., excluding leap years). The dividing curves plotted on these heatmaps represent the 99.5% free cooling threshold; regions to the right of these boundaries indicate climatic and temperature conditions where year-round more than 99.5% free cooling is achievable. One comparative observation is that evaporative-cooled systems attain more than 99.5% annual free cooling at significantly lower inlet water temperatures compared to dry-cooled systems. Specifically, evaporative-cooled liquid cooling systems achieve over 99.5% annual free cooling across all climate zones at a rack inlet water temperature of 33 °C. In contrast, dry-cooled liquid cooling systems require a much higher rack inlet temperature, 49 °C, to attain the same level of year-round free cooling across all climate zones.

The heatmaps in Figures 1c and 1d present the CPUE for both cooling configurations. The dividing curves in these heatmaps correspond to the 99.5% free-cooling thresholds. In regions to the left of these dividing curves, the CPUE decreases as the rack inlet water temperature rises across all climatic conditions, indicating a reduction in cooling energy consumption. In contrast, in regions to the right of the dividing curves, the CPUE remains nearly constant with rising rack inlet water temperatures, indicating negligible additional cooling energy savings. This is due to the fact that crossing the threshold does not yield additional free cooling hours. Based on these results, an inlet water temperature of 33 °C is identified as the global free-cooling temperature for evaporative-cooled liquid cooling systems, whereas a higher threshold of 49 °C is required for dry-cooled systems. The global free-cooling temperature is defined as the lowest rack inlet water temperature at which every climate zone, represented by the average of its cities, achieves at least 99.5% annual free cooling.

[Figures 2a–2c](#) present global maps of the annual free cooling ratio for evaporative-cooled liquid cooling systems at rack inlet water temperatures of 25 °C, 30 °C and 35 °C, respectively. Figures 2d–2i present the corresponding maps for dry-cooled liquid cooling systems at 25 °C, 30 °C, 35 °C, 40 °C, 45 °C and 50 °C, respectively. Figures 2j and 2k summarize the fraction of global land area that achieves different annual free cooling ratios for evaporative-cooled and dry-cooled systems, respectively. For dry-cooled systems (Figure 2k), while a threshold of 49 °C would enable 100% of global land area to use free cooling for 99.5% of the time, a threshold of 45 °C would still enable 77.2% of global land area to use free cooling for 99.5% of the time, and 100% of global land area to use free cooling for 90% of the time. A threshold of 40 °C would enable 71.0% of global land area to use free cooling for 99.5% of the time, and 100% of global land area to use free cooling for 87.7% of the time. A threshold of 35 °C would enable 35.5% of global land area to use free cooling for 99.5% of the time, and 70.3% of global land area to use free cooling for 90% of the time.

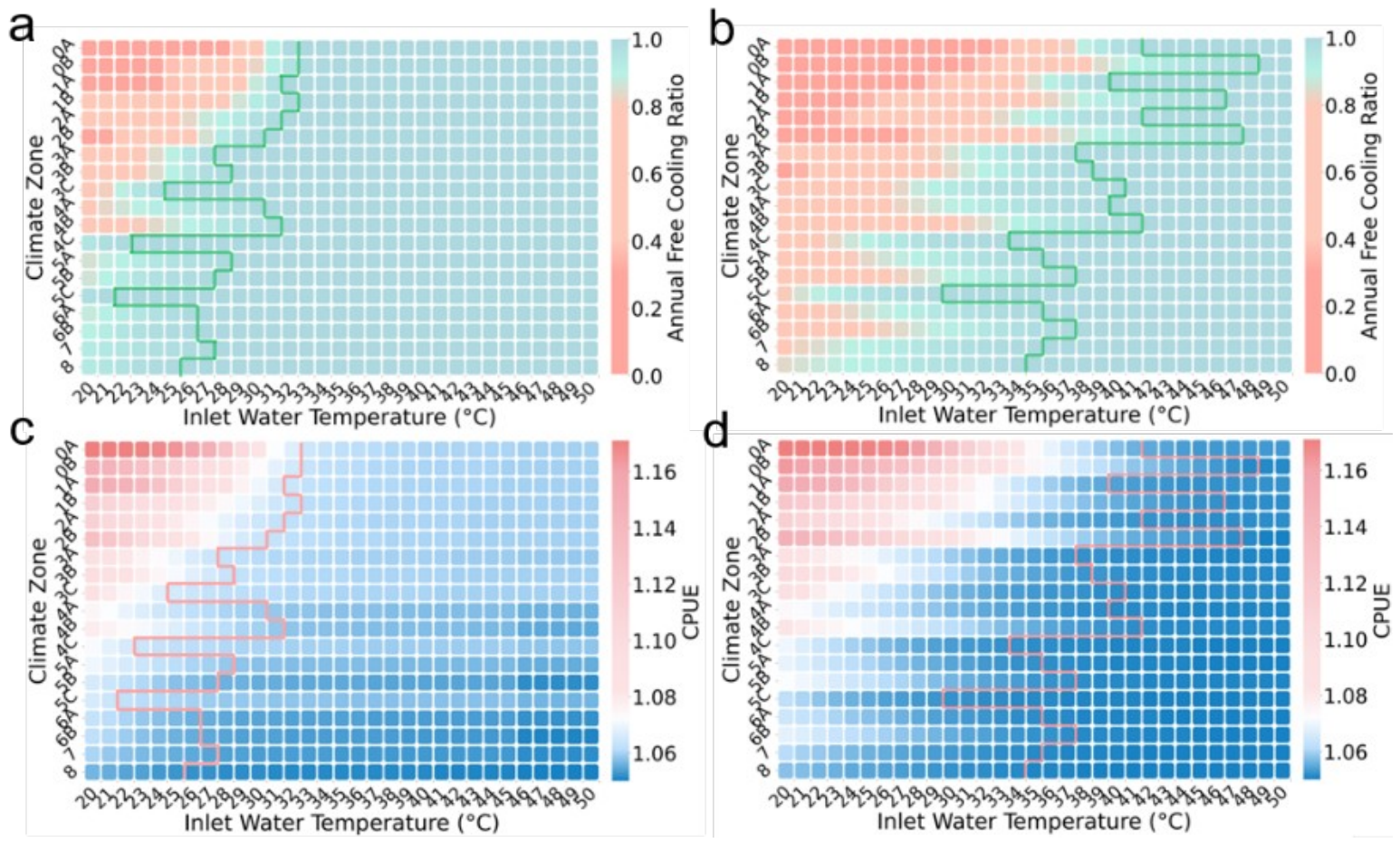

**Fig. 1**. Heatmaps of annual free cooling ratio and CPUE of dry-cooled and evaporative-cooled liquid cooling systems under different rack inlet water temperatures. **a**, Annual free cooling ratio (evaporative-cooled), **b**, Annual free cooling ratio (dry-cooled), **c**, CPUE (evaporative-cooled). **d**, CPUE (dry-cooled).

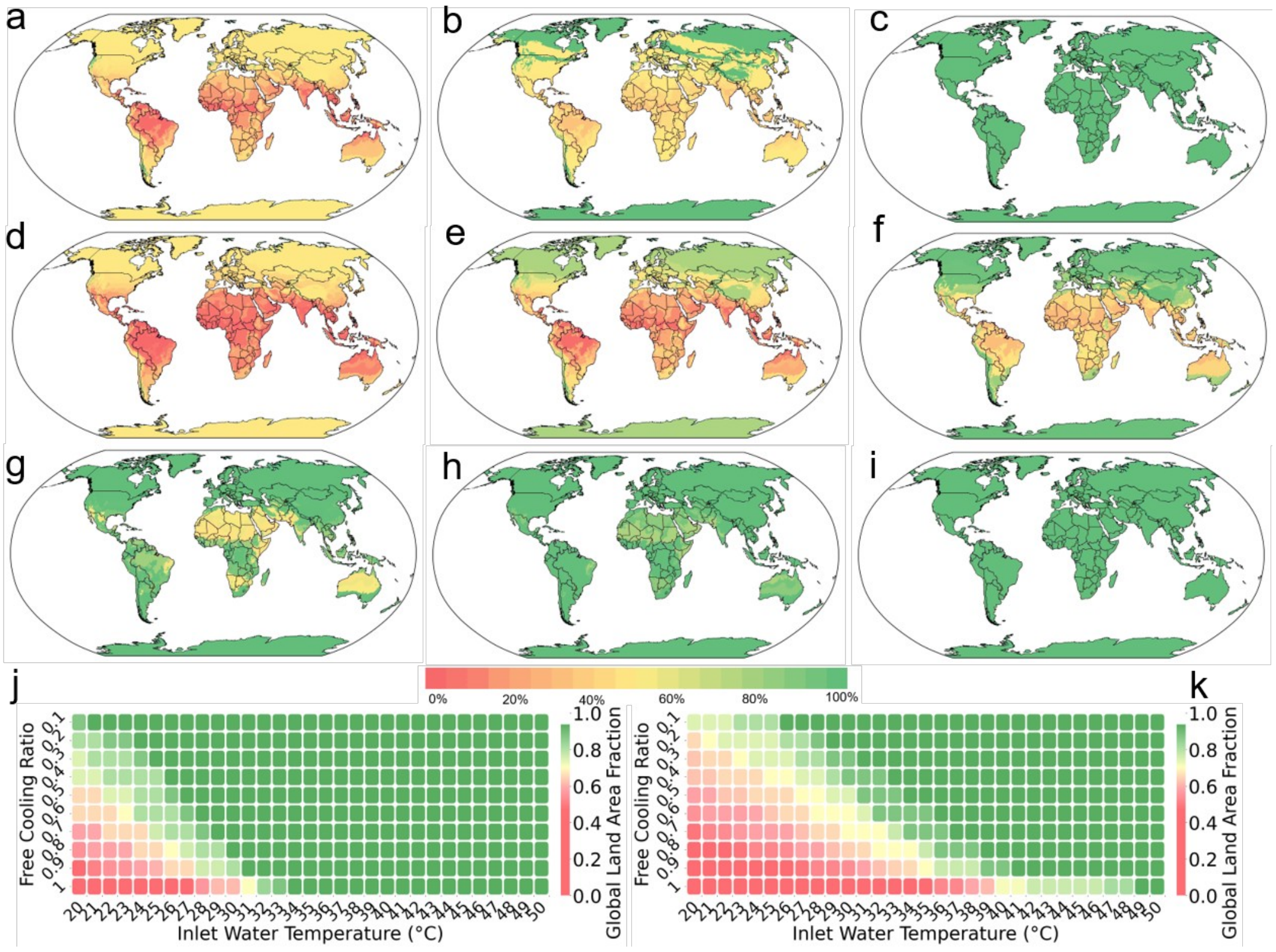

**Fig. 2**. Global map of free cooling potential of liquid cooling systems under different rack inlet water temperatures. **a**,**b**,**c**, Global maps of the annual free cooling ratio of evaporative-cooled liquid cooling systems at rack inlet water temperatures of 25 °C, 30 °C and 35 °C, respectively. **d-i,** Global maps of the annual free cooling ratio of dry-cooled liquid cooling systems at rack inlet water temperatures of 25 °C, 30 °C, 35 °C, 40 °C, 45 °C and 50 °C, respectively. **j**,**k**, Global land area fraction achieving different free cooling ratios of evaporative-cooled liquid cooling systems (j), dry-cooled liquid cooling systems (k).

**Liquid Cooling vs. Air Cooling: Energy Efficiency Improvement**

The advent of AI data centers is accelerating a paradigm shift from traditional air cooling to highly efficient liquid cooling systems for AI racks. To quantify the potential energy savings of the shift to liquid cooling, we compare evaporative and dry-cooled liquid cooling systems with the air-cooled systems commonly adopted in practice (water-cooled chiller system with waterside economizer [32]). Compared with the most recent assessment of these technologies in the United States [16], our analysis offers a finer-grained assessment of CPUE across inlet air/water temperatures and climate zones, and further quantifies the performance gap between air-cooled and liquid-cooled systems under both typical setpoints and global free cooling temperatures (100% free-cooling conditions).

Figures 3a-3b illustrate the global maps of CPUE for air-cooled data centers at a typical rack air

supply temperature of 25 °C [33] and the global free cooling rack air supply temperature of 41 °C as established in earlier work [32]. Figures 3c-3d present the global maps of CPUE for evaporative-cooled liquid cooling systems at a typical inlet water temperature of 30°C and the global free cooling rack inlet temperature of 33 °C as established in the previous section. Figures 3e-3f present the global maps of CPUE for dry-cooled liquid cooling systems at a typical inlet water temperature of 30 °C and the global free cooling rack inlet temperature of 49 °C as established in the previous section. Across all regions, both liquid cooling systems exhibit substantially lower CPUE than air cooling using a water-cooled chiller with a waterside economizer. Figure 3g provides a direct comparison of CPUE values for the aforementioned air and liquid cooling systems at their typical inlet temperature settings, 25 °C for air-cooled systems [33] and 30°C for liquid cooling systems [34]. Under typical inlet temperature settings, evaporative-cooled liquid cooling systems demonstrate a 33–56% reduction in cooling energy relative to air-cooled systems, while dry-cooled systems achieve a 25–50% reduction, depending on rack inlet water temperatures and climate conditions. Under best-practice conditions with 100% free cooling (Figure 3h), evaporative-cooled liquid cooling systems demonstrate a 22–35% reduction in cooling energy relative to air-cooled systems, while dry-cooled systems achieve a 36-39% reduction.

Furthermore, we consider recent advances in next-generation high-heat-flux liquid cooling systems, which can dissipate heat fluxes up to 3,000 $W/cm^2$ while requiring only 0.9 $W/cm^2$ using single-phase water as the coolant [12]. The proposed design achieves heat fluxes approximately three times higher than those reported in current studies in the field, while maintaining comparable pumping power. Under these conditions, the pump power required for a 10 kW IT load is approximately 2.97 W for the cold plate, which is negligible. However, the total power consumption of the secondary-side pumps must also account for pressure drops in the piping, manifolds, and other components, and is estimated to be in the range of 50-100 W [35,36]. Therefore, the CPUE of this advanced system operating under full free cooling conditions can be reasonably estimated at 1.01-1.02, using the same modelling approach as for Figures 1-3, and depending on the actual pressure drop within the liquid cooling loop. Therefore, when free cooling is fully utilized, total data center PUE can be reduced to 1.07-1.08 for current-generation evaporative and dry-cooled liquid cooling configurations and approach 1.03-1.04 for next-generation designs, assuming the deployment of 99% high efficiency UPS, and 0.5% losses for both power distribution and lighting [37] (a partial PUE of 1.02 for these subsystems).

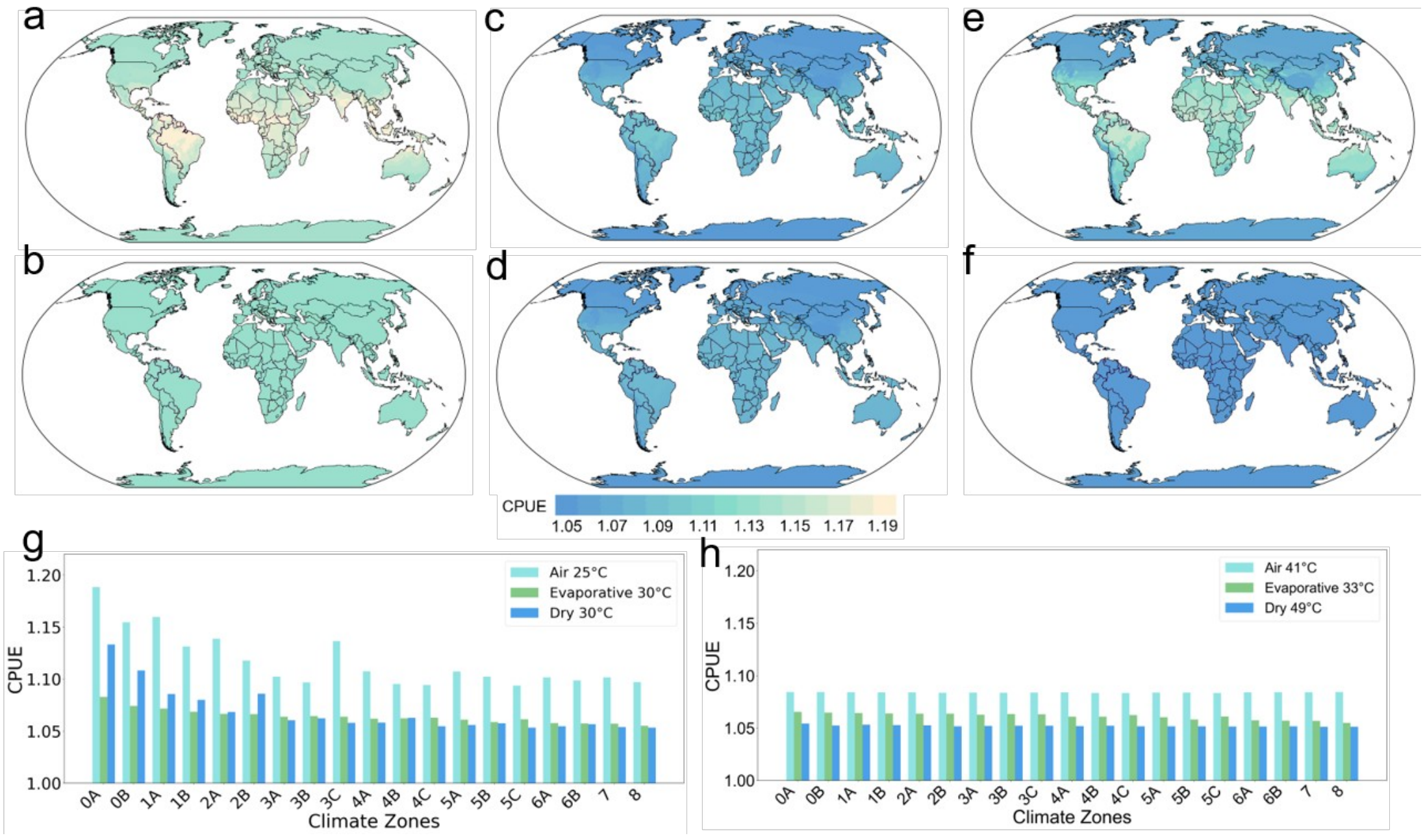

**Fig. 3**. Cooling efficiency comparison of air-cooled and liquid cooling systems. **a**,**b**, Global map of CPUE for air-cooled systems at inlet air temperatures of 25 °C and 41 °C, respectively. **c**,**d**, Global map of CPUE for evaporative-cooled liquid cooling systems at rack inlet water temperatures of 30 °C and 33 °C, respectively. **e**,**f**, Global map of CPUE for dry-cooled liquid cooling systems at rack inlet water temperatures of 30 °C and 49 °C, respectively. **g**, CPUE comparison of air-cooled systems at a representative inlet air temperature of 25 °C and liquid-cooled systems at a representative rack inlet water temperature of 30 °C. **h**, CPUE comparison of air-cooled systems at an inlet air temperature of 41 °C, evaporative-cooled liquid cooling systems at a rack inlet water temperature of 33 °C, and dry-cooled systems at 49 °C.

## Water Consumption of Liquid Cooling Systems in AI Data Centers

Beyond energy consumption, onsite water consumption has emerged as a major concern for data center cooling [23]. To enhance cooling efficiency, evaporative-cooled liquid cooling systems have been adopted in some AI data centers [38]. However, heat rejection in these systems mainly relies on the latent heat of vaporization, leading to significant water loss to the atmosphere. This water-energy nexus becomes particularly pronounced in AI facilities characterized by extreme rack densities and power scales reaching the gigawatt (GW) level [39]. To quantify this nexus, we apply a water consumption model for evaporative-cooled liquid cooling systems (see Modeling Approach). This model enables a comprehensive assessment of onsite water usage across various climatic conditions and a wide range of rack inlet water temperatures. Figure 4a presents the heatmap of Water Usage Effectiveness (WUE) [40] for evaporative-cooled liquid cooling systems in different climate zones with rack inlet water temperatures ranging from 20 °C to 50 °C. The results show that the estimated WUE ranges from 0.38 L $kWh^{-1}$ to 2.63 L $kWh^{-1}$, depending on both climate conditions and inlet water temperatures. The dividing line marks the 99.5% free cooling threshold. Prior to reaching this threshold, WUE generally decreases with higher rack inlet water temperatures. This can be attributed to the extended hours of free cooling, which minimizes the reliance on supplemental mechanical refrigeration and eliminates the additional heat of compression that would otherwise be rejected through onsite water evaporation. Beyond the free cooling threshold, WUE displays behavior that is similar to CPUE since the system no longer gains heat from compressor work, and the total heat load comes only from the servers.

Global WUE maps are presented in Figures 4b and 4c, with the corresponding climate-specific bar charts shown in Figures 4d and 4e for inlet temperatures of 30 °C and 35 °C (30°C is selected as the representative temperature for common industry practice [28], while 35°C is used for comparison to demonstrate the impact of higher inlet water temperatures).

Generally, dry-hot regions exhibit the highest water consumption, such as Climate Zone 0B. This is because, at a given dry-bulb temperature, arid regions have significantly lower relative humidity, which creates a steeper vapor pressure gradient. This increased driving force accelerates evaporation, causing these regions to rely more heavily on latent heat transfer to meet the same cooling requirements. Another observation is that regions with lower dry-bulb temperatures exhibit reduced water consumption. This is explained by the sensible heat fraction: in cooler climates, a larger portion of the thermal load is dissipated through sensible heat transfer due to the greater temperature gradient between the air and water, thereby reducing the reliance on latent heat (evaporation).

Evaporative-cooled liquid cooling systems utilize substantial water volumes for heat rejection, while dry-cooled systems dissipate heat through sensible air-to-liquid heat exchange without water consumption. A critical question arises: are evaporative-cooled systems universally more energy-efficient than their dry-cooled counterparts, indicating that to save water more cooling power must always be used [41,42]? The answer is no. Figure 4f presents the heatmap of CPUE differences between dry and evaporative liquid cooled systems across different climates and rack inlet water temperatures. Figures 4g and 4h further detail these climate-specific differences through bar charts for the 30 °C and 35 °C cases. The dividing curves in Figure 4f delineate the operational threshold between the two technologies: regions to the left of the curve indicate where evaporative-cooled systems exhibit superior energy performance (lower CPUE), whereas dry-cooled systems become slightly more efficient in the regions to the right, under the assumed control strategy of continuous spray pump operation. This shift is primarily driven by fan affinity laws, that fan power consumption increases cubically with rotational speed. Under most operating conditions, cooling tower fans function at reduced speeds, resulting in relatively low power draws that do not dominate completely the system's total energy profile. However, evaporative systems are subject to a constant energy consumption from spray pumps, which establishes a significant "energy floor." Consequently, in operational regimes where dry-cooled fans can maintain required thermal rejection at low speeds, the absence of pump-related power penalties allows dry-cooled systems to achieve superior overall efficiency compared to their evaporative counterparts.

Another key observation in Figure 4f is that evaporative-cooled systems maintain superior performance across a broader range of inlet water temperatures compared to dry-cooled systems in dry regions (Climate Zone B). This trend aligns with the water consumption data presented in Figure 4a. The performance advantage is driven by a steeper vapor pressure gradient, which facilitates increased water evaporation in dry regions and, consequently, more efficient latent heat transfer.

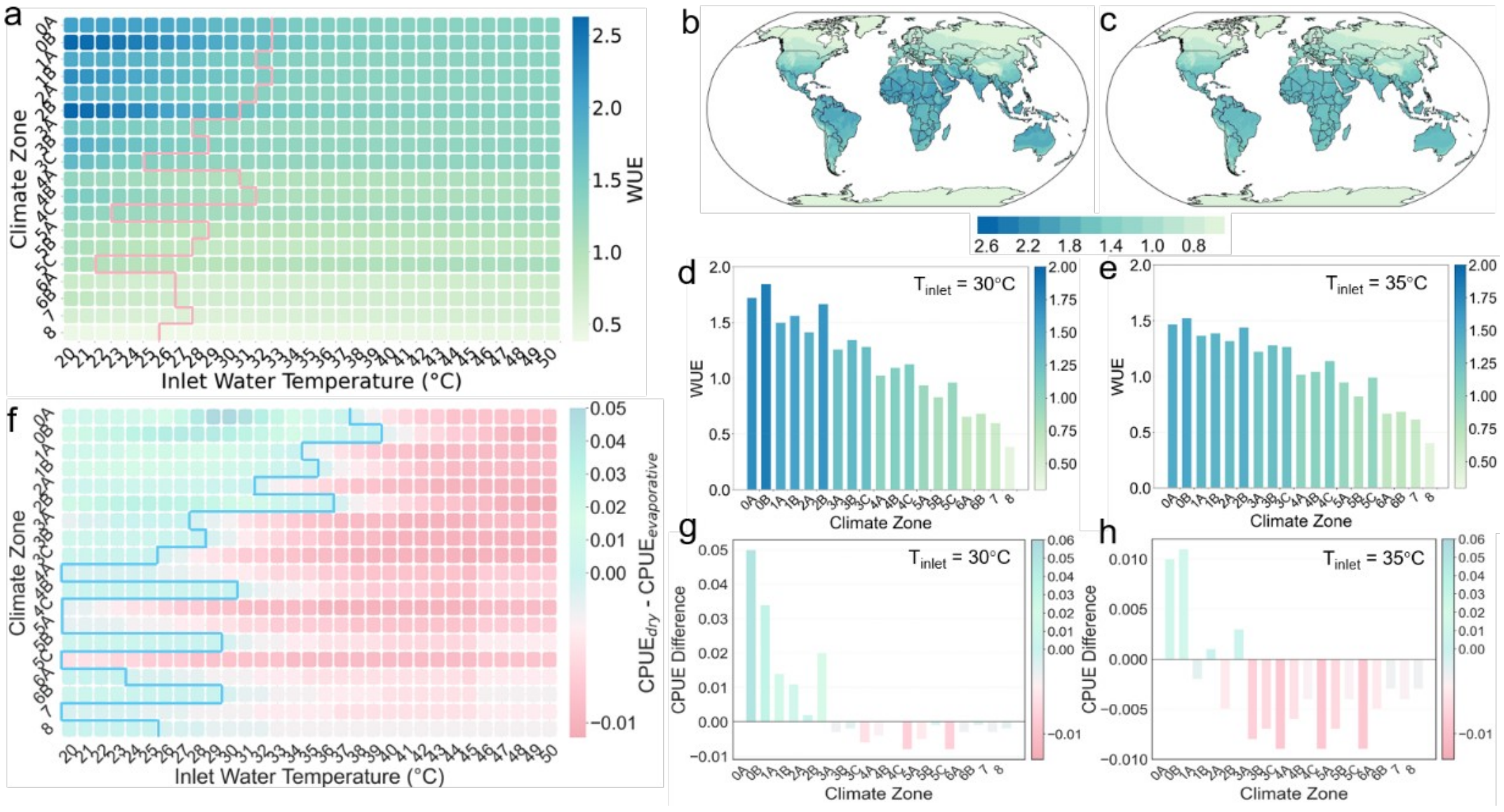

**Fig. 4**. Water usage effectiveness (WUE) of evaporative-cooled liquid cooling systems and performance comparison of evaporative-cooled and dry-cooled liquid cooling systems. **a**, Heatmap of WUE of evaporative-cooled liquid cooling systems under different climates and inlet water temperatures. **b**, **c**, Global map of WUE of evaporative-cooled liquid cooling systems at the inlet water temperature of 30 °C and 35 °C, respectively. **d**, **e**, WUE of evaporative-cooled liquid cooling systems across different climates at inlet water temperatures of 30 °C and 35 °C, respectively. **f**, Heatmap of CPUE differences between dry-cooled and evaporative-cooled liquid cooling systems under different climates and inlet water temperatures. **g**, **h**, CPUE differences between dry-cooled and evaporative-cooled liquid

cooling systems at inlet temperatures of 30 °C and 35 °C, respectively.

**Energy Trade-offs Between Server Power Increase and Cooling Energy Savings**

Elevating rack inlet water temperatures significantly reduces cooling energy consumption by extending free-cooling hours. For instance, elevating the inlet water temperatures from 20 °C to 35 °C can yield cooling energy savings of 13-60% for evaporative cooled liquid cooling systems (Figures 5a-5c) and 21-59% for dry-cooled systems (See Figure S2), depending on climate conditions. Generally, data centers in climate zones characterized by high average ambient temperatures derive greater marginal benefits from increased inlet temperatures than regions with low average ambient temperatures (wet-bulb temperatures for evaporative-cooled systems and dry-bulb temperatures for dry-cooled systems). This can be attributed to the fact that regions with low average ambient temperatures already have many free-cooling hours under baseline conditions, leaving limited additional savings from higher inlet water temperatures.

Despite the substantial cooling energy savings, higher rack inlet water temperatures may lead to server power increases. This can create a critical trade-off that dictates the total system energy efficiency. Here, we further considered the potential impact of elevated rack inlet water temperature on GPU server power and total energy consumption at the data center level. Figures 5d and 5e illustrate the total power consumption, CPUE, and GPU temperature for Hong Kong and San Francisco for evaporative cooled systems. Corresponding results for dry-cooled systems are presented in Figures 5g and 5h. The GPU power profiles are based on experimental data [43] using a machine learning training workload. Unlike synthetic stress tests that employ performance-capping or manual overclocking to reach a fixed power limit, this workload allows the GPU power draw to be governed by the hardware's inherent physical characteristics. Specifically, as junction temperatures rise with warmer coolant, increased leakage currents significantly drive-up server-level power demand. Our results show that while CPUE continues to decrease as the rack inlet water temperature increases, there is a specific inlet temperature at which the total energy consumption (i.e., server energy plus cooling energy consumption) reaches an expected minimum, which is referred to as the total energy minimum temperature. For example, the total energy minimum temperatures for Hong Kong and San Francisco are 31 °C and 21 °C, respectively when adopting evaporative-cooled liquid cooling systems. Notably, the total energy minimum temperature does not coincide with the lowest CPUE. Figure 5f and 5i summarize these total energy minimum temperatures across different climatic conditions for evaporative and dry cooled systems, respectively, where the bubble size represents the average ambient temperature of each climate zone. Generally, regions with lower average ambient temperatures exhibit lower total energy minimum temperatures. In these regions, the potential increase in server power caused by elevating the inlet temperature outweighs the potential reduction in cooling energy. This is because significant free cooling is already achieved at lower inlet temperatures, leaving limited room for further cooling energy savings. In contrast, in warmer regions, raising the rack inlet temperature significantly increases free cooling availability, leading to a net energy saving despite the increase in server power.

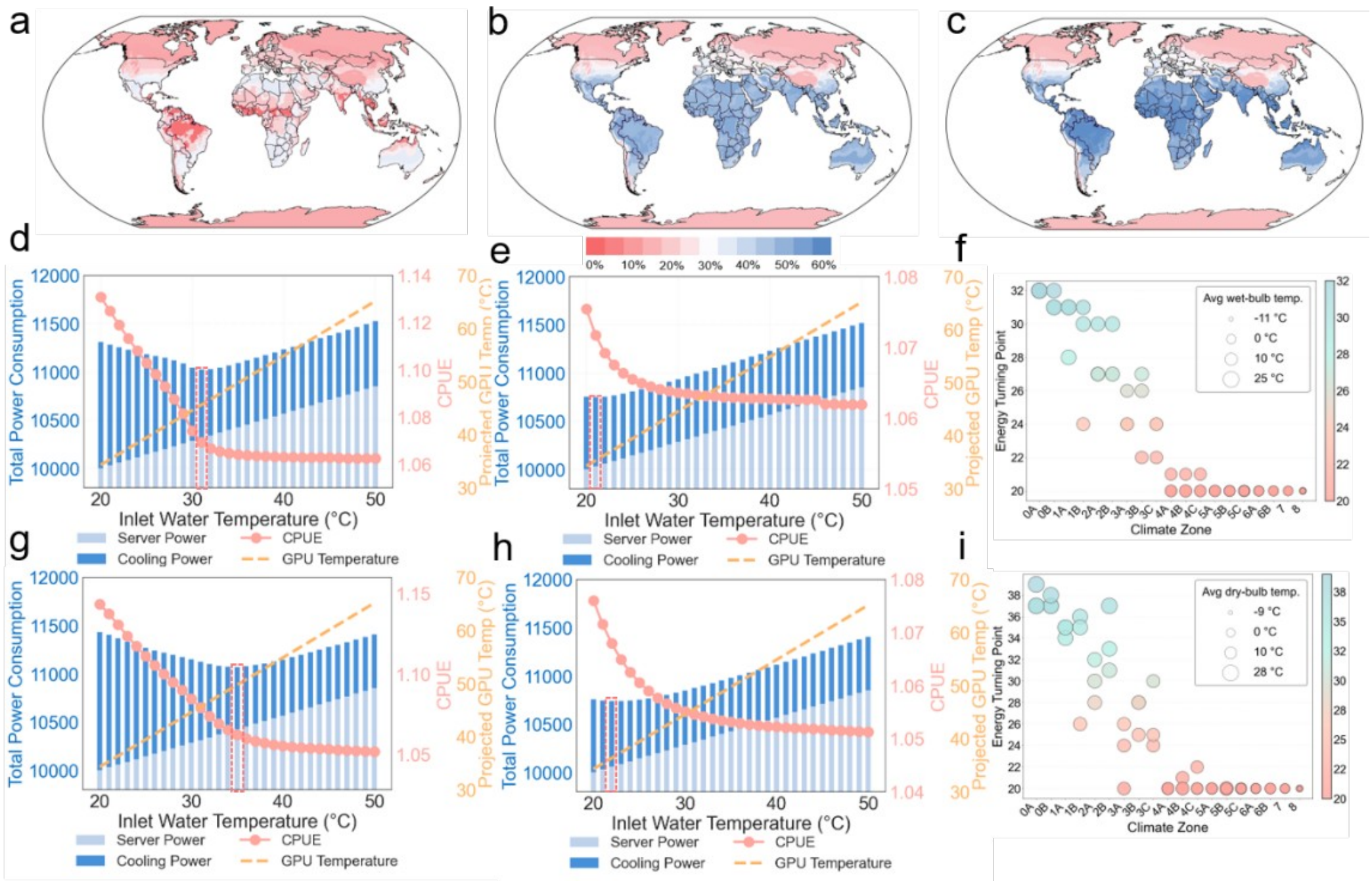

**Fig. 5**. Trade-offs between cooling energy savings and server power increase at elevated rack inlet water temperatures

**a**,**b**,**c,** Global maps of cooling energy savings for evaporative-cooled liquid cooling systems at 25 °C, 30 °C and 35 °C relative to a baseline rack inlet temperature 20 °C. **d**,**e**, Total energy performance and trade-offs between cooling energy savings and server power increase for *evaporative-cooled* liquid cooling systems in Hong Kong (**d**) and San Francisco (**e**). **f**, Total energy minimum temperatures for *evaporative-cooled* liquid cooling systems across different climate zones and their relationship with the average wet-bulb temperature. **g**,**h**, Total energy performance and the trade-off between cooling energy savings and server power increase for *dry-cooled* liquid cooling systems in Hong Kong (**g**) and San Francisco (**h**). **i**, Total energy minimum temperatures for *dry-cooled* liquid cooling systems across different climate zones and their relationship with the average dry-bulb temperature.

**Robustness and generalizability analysis**

To evaluate the robustness and generalizability of the results, we examined the variability in free-cooling thresholds, power usage effectiveness (PUE), water usage effectiveness (WUE), and cooling-energy savings across different climates and rack inlet water temperatures.

Figure 6a shows the location-level variability in the rack inlet water temperatures required to achieve a 99.5% annual free-cooling ratio. To test these climate-zone-based thresholds at the level of individual cities, each of the 57 locations was evaluated separately. A location-level analysis showed that these thresholds were sufficient to achieve a 99.5% annual free-cooling ratio in 56 of the 57 locations (98.2%). Only Abu Dhabi required higher temperatures of 34°C and 50°C for evaporative and dry cooling, respectively. These results demonstrate that the climate-zone-based thresholds are broadly applicable across climates, although slightly higher temperatures may be required under the most demanding climatic conditions. Figure 6b shows the distributions of PUE across 589 climate-zone and rack-inlet-water-temperature combinations. PUE ranged from 1.074 to 1.182 for evaporative cooling and from 1.071 to 1.191 for dry cooling, with median values of 1.083 and 1.075, respectively. Dry cooling achieved a lower PUE than evaporative cooling in 75.6% of the examined combinations. Nevertheless, the overlapping distributions and the wider range observed for dry cooling indicate that the relative performance of the two technologies remains dependent on climate and rack inlet water temperature.

Figure 6c examines the sensitivity of annual WUE to the outdoor dry-bulb temperature used to activate wet mode. At an activation temperature of 1°C, annual WUE ranged from 0.381 to 2.628 L/kWh, with a median of 1.214 L/kWh. Increasing the activation temperature to 10°C, 20°C, 30°C, and 40°C reduced the median WUE to 0.820, 0.369, 0.026, and 0.000 L/kWh, respectively. These values represent reductions of 32.4%, 69.6%, 97.9%, and 100%, respectively, relative to the 1°C case. The pronounced shift in the distributions demonstrates that WUE is highly sensitive to the wet-mode control strategy, while the variation within each distribution reflects the additional effects of climate and rack inlet water temperature. Figure 6d shows the cooling-energy savings of the evaporative and dry liquid-cooling systems at rack inlet water temperatures ranging from 20°C to 50°C across 19 climate zones. Both systems are compared with an air-cooled system using a water-cooled chiller with a waterside economizer and a typical server inlet air temperature of 25°C. The median cooling-energy savings were 42.8% for evaporative cooling and 45.2% for dry cooling and generally increased with rack inlet water temperature.

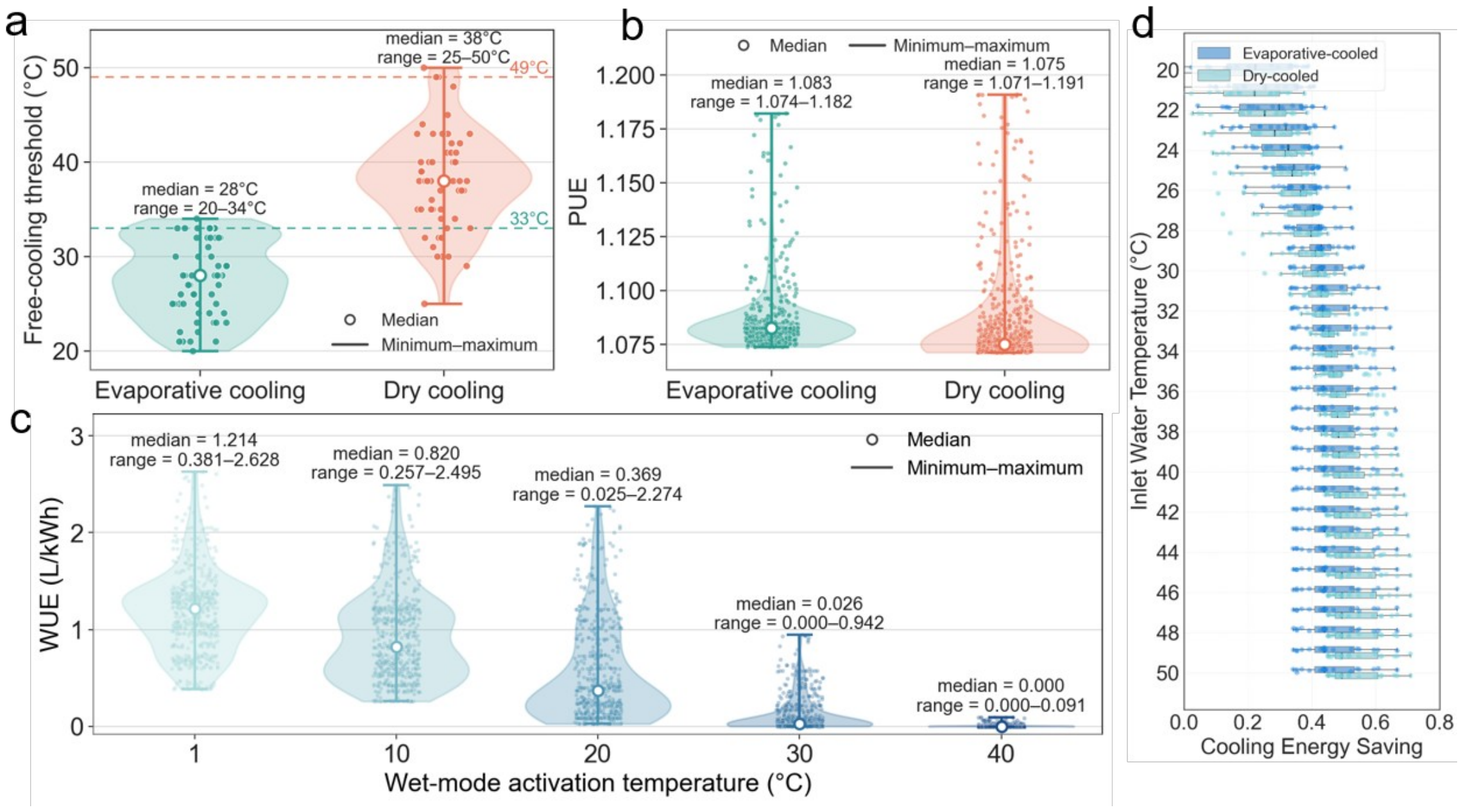

**Fig. 6**. Sensitivity of liquid-cooling performance to climate and operating conditions. **a**, Location-level variability in the rack inlet water temperatures required to achieve a 99.5% annual free-cooling ratio for evaporative and dry cooling. **b**, Distributions of PUE across 19 climate zones and rack inlet water temperatures from 20°C to 50°C. **c**, Sensitivity of annual WUE to the outdoor dry-bulb temperature used to activate wet mode. **d**, Cooling energy savings of evaporative and dry cooled liquid-cooled systems relative to air-cooled systems at a typical inlet air temperature of 25 °C under different rack inlet water temperatures.

## DISCUSSION AND POLICY IMPLICATIONS

### Total energy minimum and hardware evolution

Using current-generation GPU data, a key finding is that the minimum PUE does not necessarily coincide with the minimum total energy consumption at the system level. Although higher inlet temperatures reduce cooling energy consumption, they can also raise GPU server power consumption, resulting in a climate-dependent total energy minimum. Therefore, for data centers deploying current-generation GPU servers, designers and operators should consider the integrated optimization of both server performance and cooling efficiency rather than relying on PUE minimization alone.

However, for future GPU/TPU chip designs, this trade-off may be substantially mitigated. Improvements in chip thermal design, such as enhanced microchannel cold plates or the deployment of two-phase cooling technologies, could decouple GPU power consumption from inlet water temperature. If future chips can maintain stable power consumption as inlet water temperature increases, the total energy minimum would align with minimum cooling energy consumption. Notably, NVIDIA has indicated that its upcoming Rubin architecture may support inlet water temperatures as high as 45 °C [14], although it remains unclear whether GPU power consumption remains constant at such elevated temperatures.

Waste heat recovery opportunities

Beyond cooling optimization, higher inlet water temperatures offer significant opportunities for high-grade waste heat recovery. While there exists a globally optimal inlet temperature for maximizing free cooling and a climate-specific optimal setpoint for minimizing total data center energy use, operating at higher temperatures substantially enhances the quality and usability of recovered heat [44]. This high-grade waste heat can be reused for space heating [45], sorption-based refrigeration [46], agriculture [47], various industrial processes, such as desalination [48] and water treatment [49], and even power generation [50]. More promisingly, high-grade waste heat also presents emerging opportunities for advanced sustainability applications, such as direct air capture [51] and atmospheric water harvesting [52] in arid regions. Such heat reuse applications not only can reduce carbon emissions in downstream applications, but also improve the sustainability of data centers.

Onsite water consumption

The rapid expansion of AI and its associated data infrastructure raise critical questions regarding water sustainability. Evaporative-cooled data centers already consume substantial volumes of water for thermal management, in many cases drawing primarily from municipal water supplies [21]. Under conditions of accelerating water stress, this growing demand introduces a new and potentially significant pressure on regional water systems.

Taking China as an example, assuming that AI workloads account for approximately 50% [16] of estimated total energy consumption in 2028 (350 TWh) [2], and that these workloads are processed in AI data centers adopting evaporative-cooled liquid cooling with 100% free cooling, the total water consumption associated with AI workloads would reach approximately 192 billion liters in 2028, equivalent to the annual evaporation of about 14 West Lakes.

The world is no longer facing a temporary “water crisis,” but is instead operating under a persistent state of global water bankruptcy. Humanity is systematically liquidating its “hydrological savings”, including groundwater, glaciers, and wetlands [20]. Approximately 70% of the world’s major aquifers are in long-term decline, permanently destroying the earth's natural water storage capacity.

Taken together, these findings underscore an important implication for future AI data center development: the choice between evaporative liquid cooling and dry cooling technologies must be evaluated on a region-specific basis, with explicit consideration of energy-water trade-offs, including indirect and onsite water use, particularly in water-scarce regions. Encouragingly, the industry is trending toward higher allowable inlet water temperatures, which improves the feasibility and efficiency of dry liquid cooling systems. Looking forward, the adoption of higher inlet water temperatures combined with dry cooling architectures suggests that the AI-driven technological revolution need not impose a significant additional burden on global water resources, while still maintaining high cooling efficiency and reliable data center operation.

Policy Implications

These results carry important policy implications. First, metrics such as WUE should be integrated into data-center standards and reporting frameworks alongside PUE, with requirements tailored to different climatic conditions. The results presented can inform policymakers on the optimum performance standards for different locations, taking technology, set-point, free-cooling, and local grid and water stress, and achievable outcomes into account. Second, climate-specific guidelines for the design and operation of liquid cooling systems can help better align cooling technology choices with local climatic conditions. For instance, regions with limited municipal resources should prioritize dry or hybrid cooling systems and adopt higher cooling temperature setpoints. This helps policymakers avoid “broad brush” requirements that might not deliver maximum energy and water savings in specific locations and to identify technologies and best practices that can benefit from policy support, including financial subsidies and technical training. Finally, effective coordination in system design and operation, together with careful management of energy–water trade-offs, among regulators, utilities, and data center operators will be essential to accommodate future growth in AI computing while balancing energy demand, water sustainability, and climate resilience. The results presented here provide critical evidence based on achievable PUE and WUE outcomes, technology needs, and potential tradeoffs that will enable all stakeholders to utilize common, science-based information for discussions and decisions.

## CONCLUSIONS

This study offers a novel global analysis of how AI-driven data center cooling revolution impacts energy use, onsite water consumption, and their trade-offs. The results show that liquid-cooled systems can deliver substantially higher efficiency than air-cooled systems, demonstrating a 25-56% reduction in cooling energy relative to air-cooled systems under typical inlet temperature conditions. When free cooling is fully utilized in either liquid-cooling configuration and highly efficient UPS and power distribution systems are incorporated, PUE can be reduced to 1.07–1.08 for current-generation systems and approach 1.03–1.04 for next-generation designs.

Importantly, under continuous spray operation, evaporative-cooled systems are not always more energy-efficient than dry-cooled alternatives. In some cases, the dry-cooled configuration with elevated inlet water temperatures can achieve slightly higher cooling efficiency than the evaporative-cooled configuration while substantially reducing onsite water consumption. The selection of liquid cooling technology and the determination of inlet temperature setpoints should be evaluated in an integrated and climate-specific manner. The quantitative results presented in this study provide practical guidance for designing and operating data center cooling systems to improve energy and water efficiency. They can also inform location-specific policies and performance standards, including PUE and WUE targets, across different climate and operating conditions.

Beyond cooling system efficiency, our results reveal an important distinction between facility-level and system-level optimization: the operating condition associated with the minimum PUE does not necessarily coincide with that achieving the minimum total system energy consumption. Greater attention should therefore be given to the climate-dependent operating temperatures that minimize total system energy use, rather than relying solely on PUE minimization. This distinction is particularly important for the deployment of gigawatt-scale AI data centers, where even small improvements in system-level efficiency can translate into substantial absolute energy savings.

The rapid escalation of AI computing demand is fundamentally reshaping the energy and water requirements of data center cooling. The findings of this study provide valuable insights and practical guidance for the design and operation of liquid cooling systems in AI data centers. By explicitly incorporating both energy and water considerations into engineering decisions and policy frameworks, it is possible to support the continued growth of AI computing while addressing environmental and resource sustainability concerns.

## MODELLING APPROACH

The reference data center IT load is modelled with a representative value of 10 kW, corresponding to a single rack comprising 20 shelves as described in [53]. A typical liquid cooling system (Figure S1) model is developed for this unit based on fundamental thermodynamic principles and established component models, combined with manufacturer performance data of the cooling equipment, following the modeling framework of our previous study [32]. The specifications of individual components are determined according to the IT load (Table S2). The 10 kW unit serves as a normalized reference module: all components are sized in proportion to the IT load using fixed design ratios, such as approach temperatures, heat exchanger effectiveness, and fan and pump power per kilowatt of heat load. Because the performance relations depend on these ratios and on part-load conditions rather than on absolute equipment size, the reported metrics (power usage effectiveness, water usage effectiveness and free cooling ratio) remain applicable at larger scales. This is consistent with practice in megawatt-scale data centers, where heat rejection, heat exchange and pumping capacity are provided by multiple identical modules operating in parallel. The liquid cooling system includes chillers, heat exchangers, cooling towers, primary pumps, and secondary pumps.

**Liquid Cooling Model**

(1) Cooling tower models

Heat rejection equipment in liquid cooling systems includes closed-circuit cooling towers and dry coolers. Closed-circuit cooling towers with spray operation are selected as the representative evaporative cooling technology in this study, as they are widely used in modern data center liquid cooling systems. This configuration provides high heat rejection effectiveness through evaporative cooling while maintaining a fully closed process fluid loop, which is essential for protecting liquid-cooled IT equipment. In contrast, open-circuit cooling towers directly expose the process fluid to ambient air, which can achieve high heat transfer efficiency but may increase the risk of contamination and scaling in the system. Compared to open-circuit towers, closed-circuit systems offer better protection for sensitive IT equipment and require less maintenance, albeit sometimes with slightly lower cooling efficiency.

We develop models of closed-circuit cooling towers with evaporating water from the outside of coils containing the working fluid to cool a liquid stream. Closed-circuit cooling towers rely on the basic premise that the saturated-air enthalpy evaluated at the leaving process fluid temperature is used to represent the air–water interfacial condition. The leaving process fluid temperature under off-design conditions is determined using the following thermal-performance relation:

$$h_{sat}\left(T_{f,out}\right) = h_{a,in} + \frac{\dot{Q}_{rej}}{\dot{m}_a(1 - exp[-\beta_{des}(\frac{\dot{m}_a}{\dot{m}_{a,des}})^{y-1}])} \quad (1)$$

$$\beta_{des} = Ln[\frac{h_{sat}\left(T_{f,out,des}\right) - h_{a,in,des}}{h_{sat}\left(T_{f,out,des}\right) - h_{a,out,des}}] \quad (2)$$

where $h_{sat}\left(T_{f,out}\right)$ is the saturated-air enthalpy evaluated at the leaving process fluid temperature (J/kg); $T_{f,out}$ is the leaving process fluid temperature (°C); $h_{a,in}$ is the inlet-air enthalpy (J/kg); $\dot{Q}_{rej}$ is the heat rejection rate from the process fluid (W); $\dot{m}_a$ is the dry-air mass flow rate (kg/s); $\dot{m}_{a,des}$ is the design dry-air mass flow rate (kg/s); $\beta_{des}$ is the dimensionless design transfer parameter; and $y$ is the empirical airflow exponent (dimensionless), taken as 0.6 in this study [54]; $h_{a,out,des}$ is the design outlet-air enthalpy (J/kg); $h_{a,in,des}$ is the design inlet-air enthalpy (J/kg);

The design outlet-air enthalpy is obtained from an energy balance across the cooling tower.

$$\dot{Q}_{rej,des} = \dot{m}_{f,des} C_{p,f}\left(T_{f,in,des} - T_{f,out,des}\right) \quad (3)$$

$$h_{a,out,des} = h_{a,in,des} + \frac{\dot{Q}_{rej,des}}{\dot{m}_{a,des}} \quad (4)$$

where $\dot{Q}_{rej,des}$ is the design heat rejection rate (W); $\dot{m}_{f,des}$ is the design process fluid mass flow rate (kg/s); $c_{p,f}$ is the specific heat capacity of the process fluid (J/(kg·K)); $T_{f,in,des}$ is the design entering process fluid temperature (°C); and $T_{f,out,des}$ is the design leaving process fluid temperature (°C);

The airflow ratio under an off-design operating condition is defined by Equation (5). The airflow ratio is adjusted until the required leaving process fluid temperature is achieved, subject to the allowable fan-speed range.

$$\gamma_a = \frac{\dot{m}_a}{\dot{m}_{a,des}} \quad (5)$$

where, $\dot{m}_a$ is the dry-air mass flow rate at the operating condition (kg/s); $\dot{m}_{a,des}$ is the design dry-air mass flow rate (kg/s); and $\gamma_a$ is the airflow ratio (dimensionless).

Under ideal fan-affinity-law conditions, fan power varies with the cube of fan rotational speed and airflow rate, corresponding to k=3 in Equation (6) [55]. In this study, k=1.5 is adopted based on practical in-situ operating data, which accounts for nonideal fan performance [55].

$$\frac{P_{fan}}{P_{fan,des}} = \left(\gamma_a\right)^k \quad (6)$$

where, $P_{fan}$ is the fan power consumption of cooling towers. $P_{fan,des}$ is the fan power consumption of cooling towers at the design condition.

For dry coolers, the ε-NTU method is used to obtain the performance of the dry cooler at different conditions. Dry-bulb temperature is one of the main input parameters for the dry cooler model. Both models follow the closed-circuit cooling tower and dry fluid cooler components of TRNSYS 18, which have been widely applied in building cooling system studies. Detailed mathematical descriptions are provided in the TRNSYS 18 Mathematical Reference [54].

In the free cooling mode, the speed of the cooling tower fans is controlled to make the outlet water temperature of the heat exchangers reach the desired supply water temperatures. In mechanical cooling mode, the chilled water supply temperature is controlled by the chiller itself, and the speed of cooling tower fans is modulated to maintain the cooling tower water outlet temperature at a setpoint given by Equation (7). Where, $T_{ct,out}$ is the outlet cooling water temperature, $T_{wet}$ is the wet bulb temperature, and $T_{min,\ cd}$ is the minimum condenser water entering temperature setpoint. The cooling tower control setting given by Equation (7) is used as a proximate or near-optimal setting [56]. Cooling tower water temperature cannot drop below the chiller minimum condenser water entering temperature setpoint, as defined by the chiller manufacturer. In addition, the cooling water return temperature is limited to 45°C to avoid calcium salt precipitation [57].

$$T_{ct,out} = max(T_{wet} + 5°\text{C},\ T_{\min,\ cd}) \tag{7}$$

(2) Chiller models and heat exchanger models

Under specific weather conditions and server inlet water temperatures, when free cooling alone is insufficient to handle the entire cooling load, chillers are activated to provide supplementary cooling. For chiller models, Braun's method is used to model chillers [58], with the empirical coefficients fitted to manufacturer performance data of a high-temperature chiller; the detailed model is described in our previous study [32]. For heat exchangers [59,60], the heat transfer rate can be expressed as:

$$\dot{Q}_{hx} = \varepsilon \bullet C_{\min} \bullet (T_{h,in} - T_{c,in}) \tag{8}$$

$\dot{Q}_{hx}$ is the heat transfer rate of the heat exchanger (W); $\varepsilon$ is heat exchanger effectiveness, typically assumed to be 0.8; $C_{\min}$ is the smaller heat-capacity rate of the hot and cold streams (W/K). $T_{h,in}$ is the inlet temperature of the hot stream (°C); $T_{c,in}$ is the inlet temperature of the cold stream (°C).

(3) Pump models

The primary-side pumps are typically constant-speed pumps and are assumed to operate at their rated power. In contrast, the secondary-side pumps are variable-speed pumps. The power consumption depends on the pressure drop, water mass flow rate, and pump efficiency, as expressed in Equation (9).

$$P_{sp} = \frac{\Delta p_{sp} \times \dot{m}_w}{\rho_w \times \eta_{sp}} \tag{9}$$

$P_{sp}$ is the power consumption of the secondary-side pump (W); $\Delta p_{sp}$ is the pressure drop across the pump (Pa); $\dot{m}_w$ is the mass flow rate of the water (kg/s); $\rho_w$ is the density of water (kg/m$^3$); $\eta_{sp}$ is the pump efficiency.

**Water consumption model**

Water consumption ($\dot{V}_t$) in the closed-circuit cooling tower includes evaporation loss ($\dot{V}_{evap}$), blowdown loss ($\dot{V}_{blow}$), and drift loss ($\dot{V}_{drift}$).

$$\dot{V}_t = \dot{V}_{evap} + \dot{V}_{blow} + \dot{V}_{drift} \tag{10}$$

The heat rejection rate under each operating condition is first calculated using the actual inlet and outlet temperatures of the process fluid:

$$\dot{Q}_{rej} = \dot{m}_f c_{p,f} (T_{f,in} - T_{f,out}) \tag{11}$$

where $\dot{Q}_{rej}$ is the actual heat rejection rate (W); $\dot{m}_f$ is the process fluid mass flow rate (kg/s); $c_{p,f}$ is the specific heat capacity of the process fluid (J/(kg·K)); $T_{f,in}$ is the entering-fluid temperature (°C); $T_{f,out}$ is the actual leaving-fluid temperature (°C).

The outlet-air enthalpy is obtained from the overall energy balance [61]. The actual airflow rate ( $\dot{m}_a$) is obtained from the fan airflow ratio determined by the thermal performance model.

$$h_{a,out} = h_{a,in} + \frac{\dot{Q}_{rej}}{\dot{m}_a} \tag{12}$$

where $h_{a,in}$ is the inlet-air enthalpy (J/kg); $h_{a,out}$ is the outlet-air enthalpy (J/kg); $\dot{m}_a$ is the dry-air mass flow rate (kg/s).

For water-consumption estimation, the leaving-air state is approximated as saturated, consistent with the evaporative fluid cooler model in EnergyPlus [62] and the Merkel-based approach commonly used for cooling towers [63]. Its temperature and humidity ratio are therefore determined from the calculated outlet enthalpy, atmospheric pressure, and an outlet relative humidity of 100% [62]:

$$T_{a,out} = T(h_{a,out}, p_a, \varphi_{out} = 1) \quad (13)$$

$$\omega_{out} = \omega(T_{a,out}, p_a, \varphi_{out} = 1) \quad (14)$$

where $T_{a,out}$ is the outlet-air dry-bulb temperature (°C); $h_{a,out}$ is the outlet-air enthalpy (J/kg); $p_a$ is the atmospheric pressure (Pa); $\varphi_{out}$ is the outlet-air relative humidity (dimensionless); $\omega_{out}$ is the outlet-air humidity ratio (kg/kg).

The inlet-air enthalpy, humidity ratio, and specific volume are calculated from the measured dry-bulb temperature, relative humidity, and atmospheric pressure using standard psychrometric relationships.

The evaporation rate is calculated from the moisture balance across the air stream:

$$\dot{m}_{evap} = \dot{m}_a(\omega_{out} - \omega_{in}) \quad (15)$$

where $\dot{m}_{evap}$ is the evaporation rate (kg/s); $\dot{m}_a$ is the dry-air mass flow rate (kg/s); $\omega_{in}$ is the inlet-air humidity ratio (kg/kg); $\omega_{out}$ is the outlet-air humidity ratio (kg/kg).

Negative values of the humidity-ratio difference are set to zero because condensation is not included in the tower water-consumption estimate. Assuming a water density of approximately 1 kg/L, the hourly evaporation volume is calculated as

$$\dot{V}_{evap} = 3600\dot{m}_{evap} \quad (16)$$

where $\dot{V}_{evap}$ is the volumetric evaporation rate (L/h); $\dot{m}_{evap}$ is the evaporation rate (kg/s).

For diagnostic purposes, the latent component of heat transfer is estimated as

$$\dot{Q}_{lat} = \dot{m}_{evap} h_{fg} \quad (17)$$

where $\dot{Q}_{lat}$ is the latent heat-transfer rate (W); $\dot{m}_{evap}$ is the evaporation rate (kg/s); $h_{fg}$ is the latent heat of vaporization (J/kg).

The latent heat of vaporization is approximated as a function of the saturated outlet-air temperature:

$$h_{fg} = (2501 - 2.361 T_{a,out}) \times 10^3 \quad (18)$$

where $h_{fg}$ is the latent heat of vaporization (J/kg); $T_{a,out}$ is the saturated outlet-air temperature (°C).

The latent heat fraction ($f_{lat}$) is subsequently calculated as

$$f_{lat} = \frac{\dot{Q}_{lat}}{\dot{Q}_{rej}} \quad (19)$$

Under certain operating and ambient conditions, the entering air may undergo sensible cooling while gaining moisture through evaporation, causing the sensible heat contribution to become negative and allowing the latent heat transfer to exceed the net total heat rejected, thereby resulting in a latent heat fraction greater than unity [64,65].

The blowdown water loss, which accounts for the removal of concentrated dissolved solids, is calculated as [66,67]:

$$\dot{V}_{blow} = \dot{V}_{evap}/(CC - 1) \tag{20}$$

$\dot{V}_{blow}$ is blowdown (draw-off) water of the cooling tower (L/h); CC is the cycles of concentration, representing the ratio of dissolved solids in the circulating water to that in the make-up water. In this study, CC=9 [3,67].

The drift water loss, caused by air entrainment through the drift eliminator and air inlet, is given by [68]:

$$\dot{V}_{drift} = \frac{P_w \dot{V}_w}{100} \tag{21}$$

$\dot{V}_{drift}$ is drift water loss (L/h); $\dot{V}_w$ is the circulating water flow rate entering the cooling tower (L/h); $P_w$ is the windage loss fraction (%), typically taken as 0.01%.

To avoid overestimation of water use, the closed-circuit cooling towers are assumed to operate in dry mode with no spray water consumption when the ambient dry-bulb temperatures are at or below 0 °C. This assumption reflects standard control strategies for closed-circuit cooling towers, in which the spray system is disabled under sub-freezing conditions to prevent ice formation and ensure reliable operation.

Above 0 °C, the spray pump is assumed to operate continuously at its rated power whenever the closed-circuit cooling tower is running, which reflects a common but conservative control practice. Switching off the spray pump when dry operation can meet the load would lower both the energy use and the water use of evaporative systems; the effect of such control on water use is examined in Figure 6c.

**Server power model**

The dependence of IT power on rack inlet water temperature is derived from measurements on a cold-plate-cooled GPU server running a machine learning training workload at inlet coolant temperatures of 24 °C, 39 °C and 47 °C [43]. Over this range, the measured server power increases approximately linearly with inlet temperature. A linear fit to these data, normalized to the server power at 20 °C, gives an increase of about 0.28% per °C and is applied to the IT load at each rack inlet water temperature from 20 °C to 50 °C, with slight extrapolation beyond the measured range.

**Data center metrics**

Power Usage Effectiveness (PUE)

PUE is a widely adopted metric for evaluating the overall energy efficiency of a data center [31]. It is defined as the ratio of the total facility energy consumption to the energy consumed by IT equipment. PUE is particularly useful for benchmarking and comparing different cooling technologies and system configurations.

$$PUE = \frac{Total\ Facility\ Energy}{IT\ Equipment\ Energy} \tag{22}$$

Water Usage Effectiveness (WUE)

Water Usage Effectiveness (WUE) quantifies the amount of water consumed by a data center relative to its IT energy usage [40]. It is typically expressed as liters of water per kilowatt-hour (L/kWh) of IT load.

$$WUE = \frac{Annual\ Water\ Usage}{IT\ Equipment\ Energy} \tag{23}$$

Cooling Power Usage Effectiveness (CPUE)

CPUE is defined as a partial PUE metric used to isolate the cooling system's contribution to the overall PUE.

$$CPUE = \frac{Cooling\ Energy\ +\ IT\ Equipment\ Energy}{IT\ Equipment\ Energy} \tag{24}$$

## LIMITATIONS AND FUTURE WORK

The liquid cooling systems modeled in this study are limited to single-phase cold-plate liquid cooling systems, currently the most widely deployed liquid cooling technology in liquid-cooled data centers [44]. Therefore, the detailed quantitative conclusions are specifically applicable to this cooling approach. Advanced liquid cooling configurations [69], such as two-phase immersion cooling, are not explicitly modeled. However, the overall trends in cooling energy consumption and water use under different inlet liquid temperatures are expected to remain broadly applicable to these systems. The primary differences arise from variations in chip-level pumping power and coolant circulation-loop energy consumption [12]. In addition, the water consumption analysis in this study only considers onsite water use within the data center cooling infrastructure, without accounting for indirect water consumption associated with electricity generation [16].

Future work will focus on developing a more comprehensive and higher-fidelity modeling framework for next-generation two-phase immersion and cold-plate liquid systems that can support future ultra-high chip power densities. Such work should further evaluate the achievable PUE and WUE among these emerging cooling technologies. On the other hand, another important direction for future work is to incorporate the indirect water use associated with electricity generation. Since minimizing total water consumption may involve trade-offs between onsite water efficiency, cooling energy minimization, overall power efficiency, and the indirect water intensity of the electricity grid, a more comprehensive water-energy nexus assessment will be necessary for future sustainable data centers.

## ACKNOWLEDGMENTS

The research of Y.Z. and S.W. is financially supported by a General Research Fund (no. 152257/25E) and a grant under the Hong Kong PhD Fellowship Scheme of the Research Grants Council in the Hong Kong SAR. E.M. acknowledges funding via the Energy Demand Changes Induced by Technological and Social Innovations (EDITS) network, an initiative coordinated by the Research Institute of Innovative Technology for the Earth (RITE) and the International Institute for Applied Systems Analysis (IIASA) and funded by the Ministry of Economy, Trade and Industry (METI), Japan.

## AUTHOR CONTRIBUTIONS

Y.Z. developed the cooling system model and wrote the draft. E.M. revised and refined the paper. S.W. revised and refined the paper, acting as the main supervisor.

## DECLARATION OF INTERESTS

The authors declare no competing interests.

## REFERENCE

1. Energy and AI, International Energy Agency. 2025. https://www.iea.org/reports/energy-and-ai.
2. Electricity 2025: Analysis and forecast to 2027. International Energy Agency. https://iea.blob.core.windows.net/assets/7c671ef6-2947-4e87-beea-af0e1288e1d7/Electricity2025.pdf.
3. Lei, N., and Masanet, E. (2022). Climate- and technology-specific PUE and WUE estimations for U.S. data centers using a hybrid statistical and thermodynamics-based approach. Resources, Conservation and Recycling *182*, 106323. https://doi.org/10.1016/j.resconrec.2022.106323.
4. Fleischer, A.S. (2020). Cooling our insatiable demand for data. Science *370*, 783-784. 10.1126/science.abe5318.
5. Niemann, J., Brown, K., and Avelar, V. (2011). Impact of hot and cold aisle containment on data center temperature and efficiency. Schneider Electric Data Center Science Center, White Paper *135*, 1-14.
6. Asgari, S., MirhoseiniNejad, S., Moazamigoodarzi, H., Gupta, R., Zheng, R., and Puri, I.K. (2021). A gray-box model for real-time transient temperature predictions in data centers. Applied Thermal Engineering *185*, 116319. https://doi.org/10.1016/j.applthermaleng.2020.116319.
7. Density choices for AI training are increasingly complex. 2025. https://journal.uptimeinstitute.com/density-choices-for-ai-training-are-increasingly-complex/.
8. State of the Data Center 2026 Executive Summary. https://cdn.ymaws.com/afcom.com/resource/resmgr/resource_center/whitepapers/afcom_stateofdatacenter26_ex.pdf.

9. Vertiv unveils end-to-end AI power and cooling solutions to simplify data center infrastructure selection and deployment in EMEA. 2024. https://www.vertiv.com/en-emea/about/news-and-insights/news-releases/vertiv-unveils-end-to-end-ai-power-and-cooling-solutions-to-simplify-data-center-infrastructure-selection-and-deployment-in-emea/.
10. Nvidia's Rubin Ultra NVL576 rack expected to be 600kW, coming second half of 2027. 2025. https://www.datacenterdynamics.com/en/news/nvidias-rubin-ultra-nvl576-rack-expected-to-be-600kw-coming-second-half-of-2027/.
11. Chill Factor: NVIDIA Blackwell Platform Boosts Water Efficiency by Over 300x. 2025. https://blogs.nvidia.com/blog/blackwell-platform-water-efficiency-liquid-cooling-data-centers-ai-factories/.
12. Wu, Z., Xiao, W., He, H., Wang, W., and Song, B. (2025). Jet-enhanced manifold microchannels for cooling electronics up to a heat flux of 3,000 W cm− 2. Nature Electronics, 1-8.
13. Hu, Y., Lei, Y., Liu, X., and Yang, R. (2025). Record-high heat transfer performance of spray cooling on 3D-printed hierarchical micro/nano-structured surface. Science Bulletin *70*, 223-231. https://doi.org/10.1016/j.scib.2024.10.028.
14. TSMC x Nvidia: Breaking the Thermal Wall: How Advanced Cooling Is Powering the Future of Computing. 2025. https://tspasemiconductor.substack.com/p/tsmc-x-nvidia-breaking-the-thermal?r=4hpdhj&utm_medium=ios&triedRedirect=true.
15. Frore Systems Introduces LiquidJet™ Nexus - Defining the AI Thermal Stack for Next-Generation ½ U AI Server Systems like NVIDIA Kyber. 2026. https://www.froresystems.com/media-room/frore-systems-introduces-liquidjettm-nexus---defining-the-ai-thermal-stack-for-next-generation-1-2-u-ai-server-systems-like-nvidia-kyber.
16. Shehabi, A., Hubbard, A., Newkirk, A., Lei, N., Siddik, M.A.B., Holecek, B., Koomey, J., Masanet, E., and Sartor, D. (2024). 2024 United States Data Center Energy Usage Report.
17. Li, P., Yang, J., Islam, M.A., and Ren, S. (2025). Making AI Less 'Thirsty'. Commun. ACM *68*, 54–61. 10.1145/3724499.
18. The world's AI generators: rethinking water usage in data centers to build a more sustainable future. 2024. https://news.lenovo.com/data-centers-worlds-ai-generators-water-usage/.
19. Sustainable data center roadmap. 2025. https://icef.go.jp/wp-content/themes/icef_new/pdf/roadmap/icef2025_roadmap%20Oct%209.pdf.
20. Madani, K. (2026). Water Bankruptcy: The Formal Definition. Water Resources Management *40*, 78. 10.1007/s11269-025-04484-0.
21. Mytton, D. (2021). Data centre water consumption. npj Clean Water *4*, 11.
22. Arizona's water is drying up, that's not stopping the data center rush. 2026. https://grist.org/technology/arizona-water-data-centers-semiconducters/.
23. Lei, N., Lu, J., Shehabi, A., and Masanet, E. (2025). The water use of data center workloads: A review and assessment of key determinants. Resources, Conservation and Recycling *219*, 108310. https://doi.org/10.1016/j.resconrec.2025.108310.
24. Concern UK's AI ambitions could lead to water shortages. 2025. https://www.bbc.com/news/articles/ce85wx9jjndo.
25. ASHRAE (2021). Standard 169-2021, Climatic Data for Building Design Standards.
26. Liquid Cooling Guidelines for Datacom Equipment Centers, 2nd ed. American Society of Heating, Refrigerating and Air-Conditioning Engineers (ASHRAE). 2014.
27. Installation, Operation & Maintenance Guide InRackCDU (4U-V2). Asetek. 2019. https://manuals.plus/m/fcabf9c9d126396884fb4317b1bde28eb8f42a4afa0d3926663db063f6b5bf77.pdf.
28. Best Practice Overview Learning from inspiring projects for waste heat recovery from data centers. 2024. https://www.bytes2heat.de/static/files/20240403_BP%C3%9C_V2404_eng.pdf.
29. Meyer, N., Ries, M., Solbrig, S., and Wettig, T. (2013). iDataCool: HPC with hot-water cooling and energy reuse. (Springer), pp. 383-394.
30. https://energyplus.net/.
31. PUE: A comprehensive examination of the metric. The Green Grid. 2014. https://datacenters.lbl.gov/sites/default/files/WP49-PUE%20A%20Comprehensive%20Examination%20of%20the%20Metric_v6.pdf.
32. Zhang, Y., Li, H., and Wang, S. (2023). The global energy impact of raising the space temperature for high-temperature data centers. Cell Reports Physical Science *4*.
33. Increasing data center temperatures to reduce energy costs. Uptime Institute. 2020.

https://journal.uptimeinstitute.com/increasing-data-center-temperatures-to-reduce-energy-costs/#:~:text=Again%2C%20raising%20the%20hall%20temperatures,actual%20energy%20usage%20for%20cooling).
34. 30 °C Coolant - A durable roadmap for the future. 2024. https://ashrae.org.vn/wp-content/uploads/2024/12/30%C2%B0C-Coolant-A-Durable-Roadmap-for-the-Future-REV1_0.pdf.
35. Heydari, A., Gharaibeh, A.R., Tradat, M., Soud, Q., Manaserh, Y., Radmard, V., Eslami, B., Rodriguez, J., and Sammakia, B. (2024). Parameters of performance: A deep dive into liquid-to-air CDU assessment. Case Studies in Thermal Engineering *64*, 105461.
36. Shahi, P., Heydari, A., Modi, H., Chinthaparthy, L.S.R., Barigala, A., Eslami, B., Tradat, M., Chowdhury, U., Agonafer, D., and Rodriguez, J. (2025). Liquid to liquid in-row coolant distribution unit characterization for high power data center. Journal of Electronic Packaging *147*, 041109.
37. Lei, N., and Masanet, E. (2020). Statistical analysis for predicting location-specific data center PUE and its improvement potential. Energy *201*, 117556.
38. Supermicro Liquid Cooling Tower, https://www.supermicro.com/datasheet/datasheet_Liquid-Cooling-Tower.pdf.
39. Build times for gigawatt-scale data centers can be 2 years or less. 2025. https://epochai.substack.com/p/build-times-for-gigawatt-scale-data.
40. Water Usage Effectiveness (WUE): A Green Grid Data Center Sustainability Metric. 2011. https://airatwork.com/wp-content/uploads/The-Green-Grid-White-Paper-35-WUE-Usage-Guidelines.pdf.
41. Demonstration of Low-Cost Data Center Liquid Cooling. California Energy Commision. 2024. https://www.energy.ca.gov/sites/default/files/2024-06/CEC-500-2024-061.pdf.
42. Han, Y., Li, P., Wierman, A., and Ren, S. (2026). Small Bottle, Big Pipe: Quantifying and Addressing the Impact of Data Centers on Public Water Systems. arXiv preprint arXiv:2603.02705.
43. Ramakrishnan, B., Turner, C., Alissa, H., Trieu, D., Rivera, F., Melton, L., Rao, M., Chigullapalli, S., Getachew, T., and Prodanovic, V. (2025). Understanding the impact of data center liquid cooling on energy and performance of machine learning and artificial intelligence workloads. Journal of Electronic Packaging *147*, 021003.
44. Zhang, Y., Tang, H., Li, H., and Wang, S. (2025). Integration and interaction of next-generation AI-focused data centers with smart grids and district energy systems: The state-of-the-art, opportunities and challenges. Renewable and Sustainable Energy Reviews *224*, 116097. https://doi.org/10.1016/j.rser.2025.116097.
45. Lu, T., Lü, X., Välisuo, P., Zhang, Q., and Clements-Croome, D. (2024). Innovative approaches for deep decarbonization of data centers and building space heating networks: Modeling and comparison of novel waste heat recovery systems for liquid cooling systems. Applied Energy *357*, 122473. https://doi.org/10.1016/j.apenergy.2023.122473.
46. Pan, Q., Peng, J., and Wang, R. (2021). Application analysis of adsorption refrigeration system for solar and data center waste heat utilization. Energy Conversion and Management *228*, 113564. https://doi.org/10.1016/j.enconman.2020.113564.
47. Ljungqvist, H.M., Mattsson, L., Risberg, M., and Vesterlund, M. (2021). Data center heated greenhouses, a matter for enhanced food self-sufficiency in sub-arctic regions. Energy *215*, 119169. https://doi.org/10.1016/j.energy.2020.119169.
48. Yang, R., Konstantinou, C., and Hou, Y. (2026). Enabling grid-interactive data center-desalination coordination through thermo-electric coupling-based waste heat utilization. Applied Energy *408*, 127428. https://doi.org/10.1016/j.apenergy.2026.127428.
49. Wang, A., Zhang, C., Xu, T., Savic, D., Jiang, J., Xiao, P., He, C., Tao, Y., Daigger, G., and Ren, N. (2026). Global data–water symbiosis reduces AI infrastructure's carbon and water footprint. Environmental Science and Ecotechnology *31*, 100702. https://doi.org/10.1016/j.ese.2026.100702.
50. Corigliano, O., Algieri, A., and Fragiacomo, P. (2024). Turning Data Center Waste Heat into Energy: A Guide to Organic Rankine Cycle System Design and Performance Evaluation. Applied Sciences *14*, 6046.
51. Díaz-Marín, C.D., and Berquist, Z.J. (2025). Flipping the switch: carbon-negative and water-positive data centers through waste heat utilization. Energy & Environmental Science *18*, 8403-8413.
52. AI data centre waste heat could be used for water purification and carbon capture. 2026. https://environment.ec.europa.eu/news/ai-data-centre-waste-heat-could-be-used-water-purification-and-carbon-capture-2026-03-30_en
53. Carbó, A., Oró, E., Salom, J., Canuto, M., Macías, M., and Guitart, J. (2016). Experimental and numerical

analysis for potential heat reuse in liquid cooled data centres. Energy Conversion and Management *112*, 135-145.

54. TRNSYS 18, A transient systems simulation program, http://sel.me.wisc.edu/trnsys.
55. Lu, Y., Wang, S., Zhao, Y., and Yan, C. (2015). Renewable energy system optimization of low/zero energy buildings using single-objective and multi-objective optimization methods. Energy and Buildings *89*, 61-75.
56. Cheung, H., and Wang, S. (2019). Optimal design of data center cooling systems concerning multi-chiller system configuration and component selection for energy-efficient operation and maximized free-cooling. Renewable Energy *143*, 1717-1731. https://doi.org/10.1016/j.renene.2019.05.127.
57. Pontes, R.F.F., Yamauchi, W.M., and Silva, E.K.G. (2019). Analysis of the effect of seasonal climate changes on cooling tower efficiency, and strategies for reducing cooling tower power consumption. Applied Thermal Engineering *161*, 114148. https://doi.org/10.1016/j.applthermaleng.2019.114148.
58. Braun, J.E. (1988). Methodologies for the design and control of central cooling plants (The University of Wisconsin-Madison).
59. Lui, Y.Y. (2010). Waterside and Airside Economizers Design Considerations for Data Center Facilities. ASHRAE Transactions *116*, 98–108.
60. Taylor, S.T. (2014). How to design & control waterside economizers. ASHRAE Journal *56*, 30-36.
61. ASHRAE (2020). ASHRAE Handbook: HVAC Systems and Equipment, CHAPTER 40 COOLING TOWERS.
62. Cooling Towers and Evaporative Fluid Cooler. Engineering Reference, EnergyPlus 24.1. https://bigladdersoftware.com/epx/docs/24-1/engineering-reference/cooling-towers-and-evaporative-fluid-coolers.html.
63. Kloppers, J.C., and Kröger, D.G. (2005). A critical investigation into the heat and mass transfer analysis of counterflow wet-cooling towers. International journal of heat and mass transfer *48*, 765-777.
64. Zargar, A., Kodkani, A., Peris, A., Clare, E., Cook, J., Karupothula, P., Vickers, B., Flynn, M.R., and Secanell, M. (2022). Numerical analysis of a counter-flow wet cooling tower and its plume. International Journal of Thermofluids *14*, 100139. https://doi.org/10.1016/j.ijft.2022.100139.
65. Nasrabadi, M., and Finn, D.P. (2014). Performance analysis of a low approach low temperature direct cooling tower for high-temperature building cooling systems. Energy and Buildings *84*, 674-689. https://doi.org/10.1016/j.enbuild.2014.09.019.
66. Lei, N., and Masanet, E. (2022). Climate-and technology-specific PUE and WUE estimations for US data centers using a hybrid statistical and thermodynamics-based approach. Resources, Conservation and Recycling *182*, 106323.
67. Xia, L., Li, J., Ma, W., Gurgenci, H., Guan, Z., and Wang, P. (2017). Water consumption comparison between a natural draft wet cooling tower and a natural draft hybrid cooling tower—An annual simulation for Luoyang conditions. Heat Transfer Engineering *38*, 1034-1043.
68. Ministry of Housing and Urban-Rural Development of the People's Republic of China, 2016. Code for design of cooling tower for mechanical ventilation (GB/T 50392–2016). China Planning Press, Beijing.
69. Chen, X., Hu, C., Liang, J., Li, J., Lu, L., Wu, Y., and Zang, J. (2025). Efficient two-phase liquid cooling and optimization technology on the chip side based on high-power data centers: Summary and prospects. Renewable and Sustainable Energy Reviews *223*, 115994. https://doi.org/10.1016/j.rser.2025.115994.

## Supplemental Information

### Supplemental Table

Table S1. Climate zones [1] and 57 representative cities

| Zones | Description | Representative Cities | | |
|---|---|---|---|---|
| **0A** | Extremely Hot Humid | Bangkok (Thailand) | Singapore | Manila (Philippines) |
| **0B** | Extremely Hot Dry | Abu Dhabi (United Arab Emirates) | Guantanamo Bay (Cuba) | Ahmedabad (India) |
| **1A** | Very Hot Humid | Hong Kong (China) | Campina Grande (Brazil) | Veracruz (Mexico) |
| **1B** | Very Hot Dry | New Delhi (India) | Richmond (Australia) | Yuanjiang (China) |
| **2A** | Hot Humid | Fuzhou (China) | Sao Paulo (Brazil) | Tunis (Tunisia) |
| **2B** | Hot Dry | Cairo (Egypt) | Amritsar (India) | Makindu (Kenya) |
| **3A** | Warm Humid | Guiyang (China) | Istanbul (Turkey) | Roma Ciampino (Italy) |
| **3B** | Warm Dry | Adelaide (Australia) | Cozzo Spadaro (Italy) | Melilla (Spain) |
| **3C** | Warm Marine | Kunming (China) | Katanning (Australia) | Santander (Spain) |
| **4A** | Mixed Humid | Beijing (China) | Brussels (Belgium) | Lyon (France) |
| **4B** | Mixed Dry | Madrid (Spain) | Tianjin (China) | AnYang (China) |
| **4C** | Mixed Marine | Seattle (United States) | Vancouver (Canada) | San Francisco (United States) |
| **5A** | Cool Humid | Yingkou (China) | Birmingham (UK) | Hamburg (Germany) |
| **5B** | Cool Dry | Taiyuan (China) | Cheyenne (United States) | Lanzhou (China) |
| **5C** | Cool Marine | Comox (Canada) | Sandspit (Canada) | Ketchikan (United States) |
| **6A** | Cold Humid | Changchun (China) | Moscow (Russia) | Gander (Canada) |
| **6B** | Cold Dry | Chifeng (China) | Alamosa (United States) | Datong (China) |
| **7** | Very Cold | Harbin (China) | Anchorage (United States) | Ostersund (Sweden) |
| **8** | Subarctic/Arctic | Mohe (China) | Yakutsk (Russia) | Churchill (Canada) |

Table S2. Specifications of the liquid cooling system components [2-4]

| Equipment | Parameter | Value |
|---|---|---|
| Secondary pump | Volumetric flow rate (m3/h) | 2.1 |
| | Rated power (kW) | 0.252 |
| Heat exchanger | Effectiveness | 0.8 |
| Primary pump | Volumetric flow rate (m3/h) | 2.1 |
| | Rated power (kW) | 0.252 |
| Dry cooler | Design inlet fluid temperature (°C) | 45 |
| | Design outlet fluid temperature (°C) | 40 |
| | Specific heat capacity (kJ/kg·°C) | 3.5 (water with propylene glycol) |
| | Design air flow rate (kg/h) | 33768 |
| | Design fluid flow rate (kg/h) | 5292 |
| | Rated fan power (kW) | 1.76 |
| Closed-circuit cooling tower | Design air flow rate (kg/h) | 20227 |
| | Design fluid flow rate (kg/h) | 5554 |
| | Design inlet fluid temperature (°C) | 34 |
| | Design outlet fluid temperature (°C) | 29 |
| | Rated fan power (kW) | 1.1 |
| Spray pump | Rated power (kW) | 0.25 |

Note: The fan and pump specifications of the closed-circuit cooling tower system follow a 10 kW cold-plate liquid cooling test bench [2]. The spray pump power was recalibrated from 0.75 kW to 0.25 kW, because the test-bench

spray pump is oversized relative to its heat load [4]. The dry cooler specifications follow a commercial unit used in a comparable modeling study [3]. In this study, both heat rejection units are oversized relative to the 10 kW IT load, each with a nominal heat rejection capacity of about 25 kW under its rating conditions. Such oversizing is consistent with data center practice, where heat rejection capacity includes redundancy and a margin for the hottest outdoor conditions. This design allows smaller approach temperatures and lower fan speeds during operation [2], and the comparable capacities ensure a fair comparison between the two technologies.

## Supplemental Figures

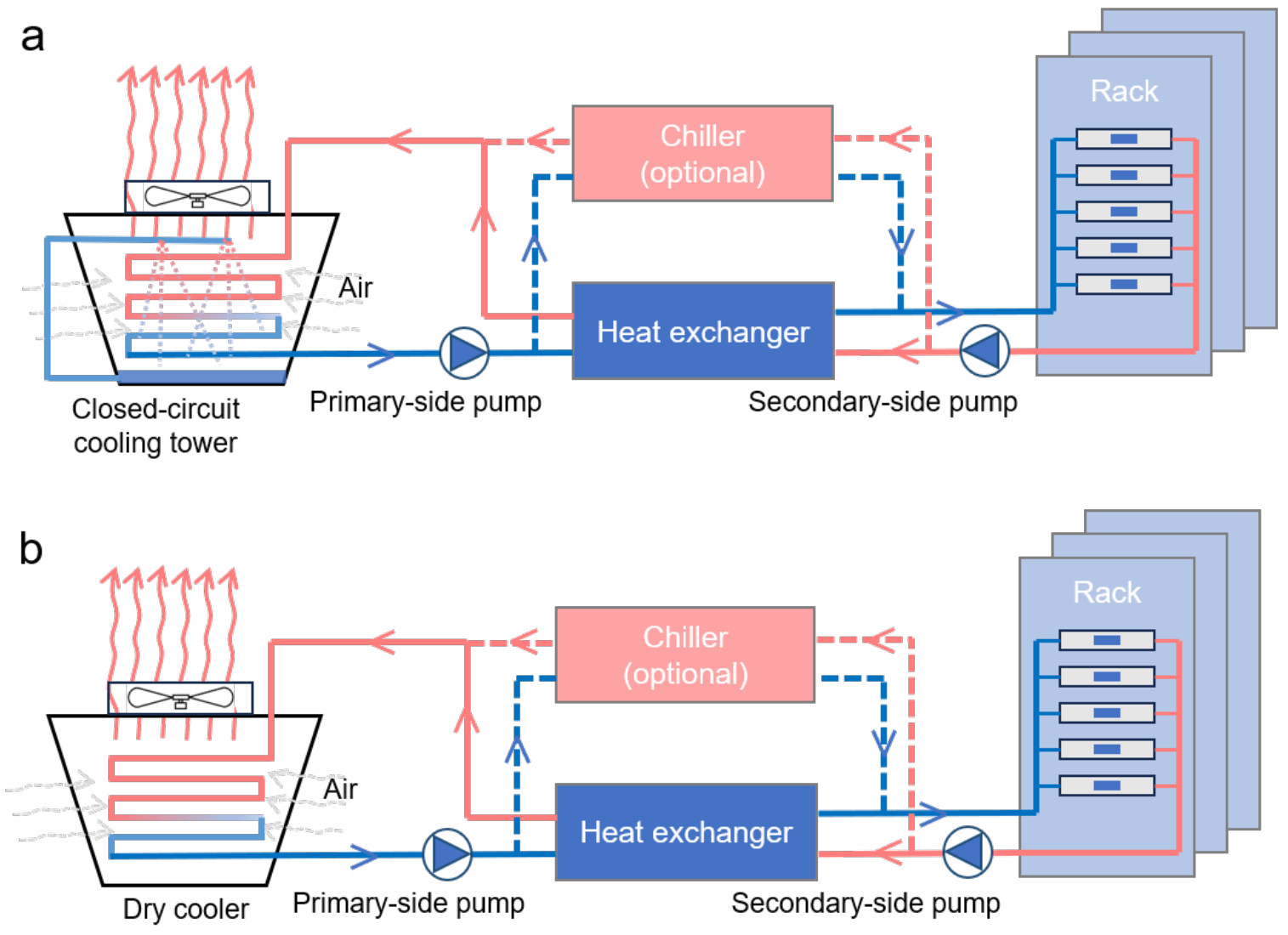

**Figure S1**. Schematics of the evaporative-cooled (**a**) and dry-cooled (**b**) cold-plate liquid cooling systems
Note: In free cooling mode, the chiller is off, and all heat is rejected by the closed-circuit cooling tower (a) or the dry cooler (b) through the heat exchanger. Dashed lines indicate pipes used only when the chiller operates.

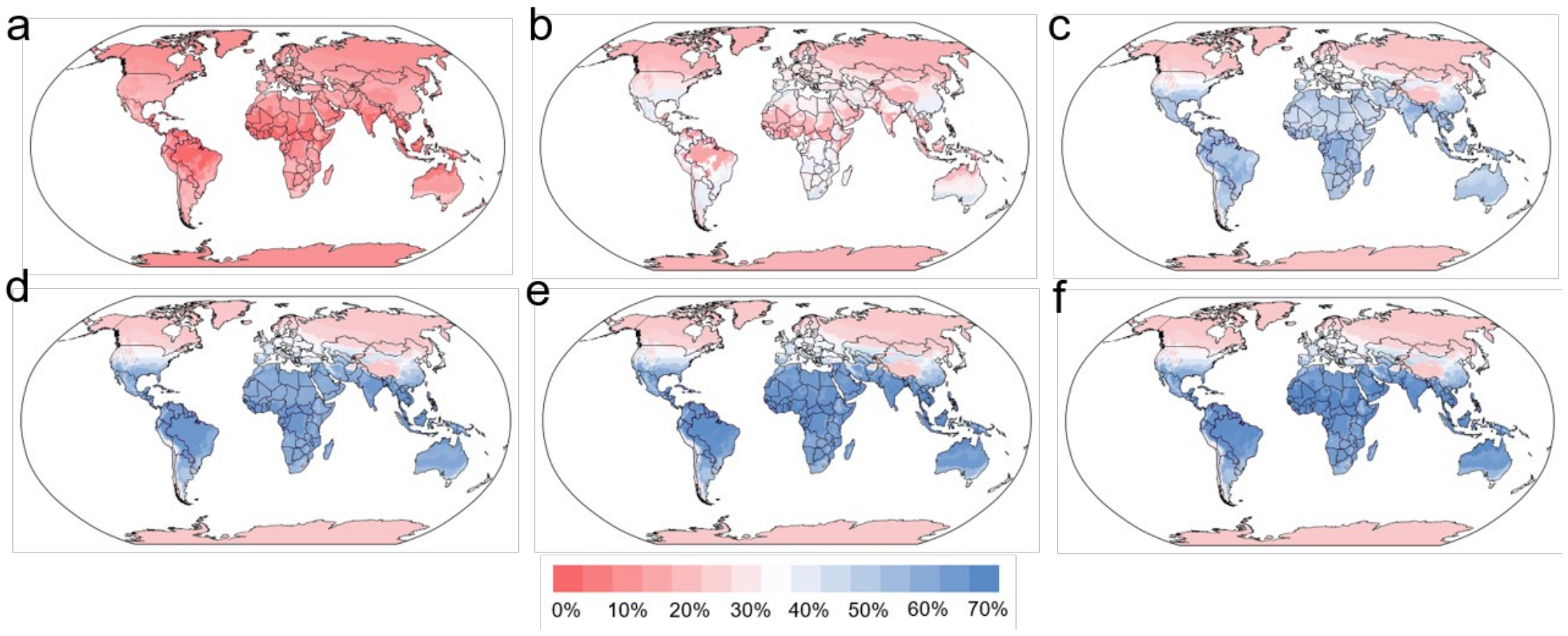

**Figure S2**. Global map of cooling energy savings of dry-cooled liquid cooling systems under different inlet water temperatures. **a-f**, Global maps of cooling energy savings of 25 °C, 30 °C, 35 °C, 40 °C, 45 °C and 50 °C relative to a baseline inlet water temperature 20 °C, respectively.

## Supplemental References

1. ASHRAE (2021). Standard 169-2021, Climatic Data for Building Design Standards.
2. Qu, S., Duan, K., Guo, Y., Feng, Y., Wang, C., and Xing, Z. (2024). Real-time optimization of the liquid-cooled data center based on cold plates under different ambient temperatures and thermal loads. Applied Energy *363*, 123101.
3. Lu, T., Lü, X., Välisuo, P., Zhang, Q., and Clements-Croome, D. (2024). Innovative approaches for deep decarbonization of data centers and building space heating networks: Modeling and comparison of novel waste heat recovery systems for liquid cooling systems. Applied Energy *357*, 122473.
4. https://baltimoreaircoil.com/products/closed-circuit-cooling-towers/series-v-closed-circuit-cooling-tower.